\documentclass[twocolumn]{aastex62}

\usepackage{graphics,epsf}
\usepackage{amsmath}                
\usepackage{amsfonts}               
\usepackage{amssymb}                
\usepackage{epsfig}   
\usepackage{subfigure}
\usepackage{graphicx}
\usepackage{float}
\usepackage{color}
\usepackage[para,online,flushleft]{threeparttable}

\def \cm{~\rm{cm}}
\def \s{~\rm{s}}
\def \km{~\rm{km}}

\def \K{~\rm{K}}
\def \g{~\rm{g}}

\def \erg{~\rm{erg}}

\def \yrs{~\rm{yrs}}
\def \yr{~\rm{yr}}

\begin{document}

\title{Shaping `Ears' in planetary nebulae by early jets}


\author{Muhammad Akashi}
\affiliation{Department of Physics, Technion, Haifa, 3200003, Israel; akashi@physics.technion.ac.il; soker@physics.technion.ac.il}
\affiliation{Kinneret College on the Sea of Galilee, Samakh 15132, Israel}

\author[0000-0003-0375-8987]{Noam Soker}
\affiliation{Department of Physics, Technion, Haifa, 3200003, Israel; akashi@physics.technion.ac.il; soker@physics.technion.ac.il}
\affiliation{Guangdong Technion Israel Institute of Technology, Shantou 515069, Guangdong Province, China}


\begin{abstract}
We conduct three dimensional hydrodynamical numerical simulations of planetary nebula (PN) shaping and show that jets that precede the ejection of the main PN shell can form the morphological feature of \textit{ears}. Ears are two opposite protrusions from the main nebula that are smaller than the main nebula and with a cross section that decreases monotonically from the base of an ear at the shell to its far end. Only a very small fraction of PNe has ears. The short-lived jets, about a year in the present simulations, interact with the regular asymptotic giant branch (AGB) wind to form the ears, while the later blown dense wind forms the main PN dense shell. Namely, the jets are older than the main PN shell. We also find that for the jets to inflate ears they cannot be too energetic, cannot be too wide, and cannot be too slow. A flow structure where short-lived jets precede the main phase of nebula ejection by a few years or less can result from a system that enters a common envelope evolution. The low mass companion accretes mass through an accretion disk and launches jets just before it enters the envelope of the giant progenitor star of the PN. Shortly after that the companion enters the envelope and spirals-in to eject the envelope that forms the main PN shell. 
\end{abstract}


\section{INTRODUCTION}
\label{sec:intro}
 
Many observations and theoretical studies over the years, and more so in the last decade, attribute major roles to jets in shaping planetary nebulae (PNe; e.g., 
\citealt{Morris1987, Soker1990AJ, SahaiTrauger1998, Boffinetal2012, Miszalskietal2013, Tocknelletal2014, Huangetal2016, Sahaietal2016, RechyGarciaetal2017, GarciaSeguraetal2016, Dopitaetal2018, Fangetal2018, KameswaraRaoetal2018, Lagadec2018, AliDopita2019, Derlopaeta2019, Jonesetal2019jets, Miszalskietal2019, Oroszetal2019, Scibellietal2019, Guerreroetal2020,  MonrealIberoetal2020, RechyGarciaetal2020, Soker2020Galax, Tafoyaetal2020, Zouetal2020, Guerreroetal2021}, for a small fraction of many more papers). Observations show a link between the presence of a binary central star and shaping by jets (e.g., \citealt{Boffinetal2012, Miszalskietal2013, Miszalskietal2018a}).
This link includes also post-asymptotic giant branch (AGB) stars that might not form a PN (e.g., \citealt{Thomasetal2013, Bollenetal2017, VanWinckel2017, Bollenetal2020, Bollenetal2021}). 
 
We clarify that we refer to any bipolar outflow, i.e., two opposite polar outflows with a mirror symmetry about the equatorial plane, as jets. The jets might be narrow, or the half opening angle of each jet might be large, even close to $90^\circ$. As well, the outflow in the jets might be continuous, periodic, or stochastic. We still refer to the polar outflow as a jet.  
  
The main aim of hydrodynamical simulations of jets in PNe is to show that jets can account for the different morphological features 
(e.g.,  \citealt{LeeSahai2004, Dennisetal2009, Leeetal2009, HuarteEspinosaetal2012, Balicketal2013, Akashietal2015, Balicketal2017, Akashietal2018, Balicketal2018, EstrellaTrujilloetal2019, RechyGarciaetal2019, Balicketal2020}). These and many other simulations have shown that jets can account for a very rich varieties of morphologies. 

One of the key advantages of jets is that they allow to make use of the energy source that results from mass accretion onto the companion. They introduce axially-symmetric flows that can affect the descendant nebula in many ways, depending, among others, on the intensity of the jets, their duration, and when their activity phase takes place. In this study we consider the jets to be weak, of short duration, and to take place before the main nebula ejection. 

Our present goal is to show that jets can form `Ears' in elliptical PNe. By ears we refer to two opposite protrusions from the main PN shell. Ears differ from bipolar lobes by three main properties. (1) Ears are  smaller than the main inner shell from which they protrude. Most bipolar lobes are larger than the inner main shell. (2) An ear cross section (perpendicular to the symmetry axis) monotonically decreases outward, i.e., as we move from its base at the main inner shell to its tip. Most Bipolar lobes, on the other hand, widen first, and then get narrower toward their tip. 
 A third criterion distinguishes ears from elliptical PNe. (3) The boundary between the ears and the main nebula has a dimple (two inflection points) on each side of each ear.  
Like bipolar lobes, in most cases ears are along the symmetry axis of the nebula and have different emission properties or brightness, like being fainter, than the main PN shell. 
By their definition, ears exist only in elliptical PNe. In that regards and related to our study, we assume that all elliptical PNe are shaped by binary interaction, mainly by a low mass main sequence companion that enters a common envelope evolution (CEE) with an AGB star (e.g., \citealt{Soker1997Rev}).  

We list the 10 best examples we could find of PNe with ears. We give one or two sources for the image of each PN.  
The images of the first two PNe with ears are from HASH (the Hong Kong/AAO/Strasbourg H$\alpha$ planetary nebula database; \citealt{PArkeretal2016}). In  K~3-24 we identify the ears protruding to the north and to the south, while in IC~289 (also \citealt{Hajianetal1997}) the ears are to the north-west and to the south-east, and they are not exactly aligned with the central star. 
  
The PN K~3-4 (\citealt{Manchadoetal1996}) is an interesting case. Firstly, the two ears are not aligned with the center of the PN, as in IC~289. Secondly, the ears are large, and just on the border between being lobes and being ears because their width (cross section) stays constant for some distance above their base. Since their length as projected on the plane of the sky is shorter than the main shell, we term them ears (or border-ears). 

In M 2-53 \citep{Manchadoetal1996} we identify large ears, one in the west and one in the east. The PN NGC~6905 (\citealt{Balick1987, PhillipsRamosLarios2010}) has elongated ears. We term them ears because their width (cross section) decreases monotonically to their tips. The PN NGC~3242 has two pairs of ears along the same axis (\citealt{Schwarzetal1992}). The PN NGC~6563 (\citealt{Schwarzetal1992}) has point-symmetric ears in an `S' shape. Other PN with ears are NGC 6852 \citep{Manchadoetal1996}, Na~1 \citep{Manchadoetal1996}, and M~2-40 \citep{Manchadoetal1996}.
 
The formation of ears in PNe might have relations to ears in some remnants of type Ia supernovae (SNe Ia). Most possibly is that some of these SNe Ia exploded inside a PN, i.e., a SN inside a PN (SNIP). We take the view that in remnants of SNe Ia, like in PNe, the ears are features along the polar (symmetry) axis (e.g., \citealt{TsebrenkoSoker2013}), rather than an equatorial dense gas (e.g., \citealt{Chiotellisetal2020}). In that respect we note that \cite{Blondinetal1996} form ears in type II supernovae by assuming a circumstellar gas with a high equatorial density into which the star explodes. They obtain polar ears, but not by the action of jets. 

In section \ref{sec:numerical} we describe the three-dimensional (3D) simulations and in section \ref{sec:results} we describe our results of 17 different simulations. We do not try to fit any PN particularly, but only to derive the general structure of ears, because the parameter space (jets' properties, shell properties) is very large. In section \ref{sec:Evolution} we show the evolution with time. We summarise our results in  section \ref{sec:summary}. 
       
\section{NUMERICAL SET-UP}
 \label{sec:numerical}

\subsection{The numerical scheme and the jets}
 \label{subsec:Jets}

We use version 4.2.2 of the hydrodynamical FLASH code \citep{Fryxell2000} with the unsplit PPM (piecewise-parabolic method) solver to perform our 3D hydrodynamical simulations. FLASH is an adaptive-mesh refinement (AMR) modular code used for solving hydrodynamics and magnetohydrodynamics problems. We do not include radiative cooling in the simulations because the interaction takes place in a dense region close to the binary system, such that some zones are optically thin while others are not. The inclusion of radiative transfer in this 3D complicated flow is too demanding. We instead vary the values of the adiabatic index $\gamma$. 

We employ a full 3D AMR (7 levels; $2^{9}$ cells in each direction) using a Cartesian grid $(x,y,z)$ with outflow boundary conditions at all boundary surfaces. We take the $z=0$ plane to be in the equatorial plane of the binary system, which is also the equatorial plane of the nebula. We simulate the whole space (the two sides of the equatorial plane).

In most simulations the size of the grid is $(4\times 10^{16}\cm)^{3}$. In two simulations we take twice as large a grid to follow the evolution to later times.
At time $t=0$ we fill the grid with a spherical wind with velocity of
$v_{\rm AGB}= 20 \km \s^{-1}$ and a mass loss rate $\dot M_{\rm AGB}=10^{-6} M_\odot ~{\rm yr}^{-1}$. We term this wind a regular AGB wind. 

We launch the two opposite jets from the inner $4\times 10^{14} \cm$ zone along the $z$-axis (at $x=y=0$) and  within a half opening angle of $\alpha_{\rm j}$. We chose two values of $\alpha_{\rm j}$, one represents narrow jets, as observed in many young stellar objects, and one represents wide jets, as observed in some post-AGB binary systems (e.g., \citealt{Bollenetal2021}). The injection temperature of the jets is
$10^4 \K$, a typical temperature of warm gas. The jets are active during the time period from $t=0$ to $t_{\rm j}$ and the ejection of the dense spherical shell starts one year after $t_{\rm j}$. These time scales are comparable to the dynamical time of the CEE, which we assume is the timescale during which the companion enters the envelope. 
The jets' initial velocity is $v_{\rm j} = 100$ or $v_{\rm j} = 200 \km \s^{-1}$, which is about the escape velocity from a low-mass main sequence star or from a brown dwarf (the companion star). The mass-loss rate into the two jets together is $\dot M_{\rm 2j} \simeq 10^{-4} - 10^{-5} M_\odot \yr^{-1}$. These mass loss rates are about $0.01-0.1$ times the rates that \cite{Shiberetal2019} take. 
The reasons for the lower values here are that we take a lower mass companion and that the giant in our case is a more extended AGB star with a lower envelope density compared with the red giant branch model of \cite{Shiberetal2019}. 
For numerical reasons (to avoid very low densities) we inject a very weak slow wind in the directions where we do not launch the jets, i.e., in the sector $\alpha_{\rm j}<\theta<90^\circ$ in each hemisphere (for more numerical details see \citealt{AkashiSoker2013}).

\subsection{The spherical dense shell}
 \label{subsec:shell}

In most of our previous studies (e.g., \citealt{AkashiSoker2013, Akashietal2018, AkashiSoker2018}) we injected the jets into a dense spherical shell (formed by an intensive wind), which itself was embedded in a much less dense wind (formed by the regular AGB wind). Namely, the jets active phase follows the high mass loss rate that formed the dense shell (jets are younger than the dense PN shell). Such interactions can form large bipolar lobes with different properties. 

In this study we have simulated about twenty different cases where we launched jets into a dense shell. We failed to obtain ears. Namely, we could not form  polar lobes that are smaller than the dense shell and that have a cross section that decreases with distance from the center (for definition of ears see section \ref{sec:intro}). These failures led us to conduct simulations where we launch the dense shell after we launch the jets. Such a case might be, for example, when the companion accretes mass from the AGB progenitor of the PN and launches jets. Later it enters a common envelope evolution, a process that ejects the dense shell. The jets, therefore, interact with the less-dense (regular AGB) wind that preceded the ejection of the dense shell.  
 
We assume that the main sequence companion that launches the relatively weak jets is of low mass $M_2 \simeq 0.1-0.3 M_\odot$ (in most cases; might even be a brown dwarf), and therefore after it enters the CEE it ejects an elliptical nebula rather than a bipolar nebula or a dense equatorial torus \citep{Soker1997Rev}. This assumption is compatible with the observation that ears are present mainly in elliptical PNe. In assuming that a low mass companion ejects the elliptical shell we have in mind the PN A30 that has an almost spherical morphology (not including the central knots) and has a central binary system with an orbital period of 1.06 days \citep{Jacobyetal2020}. However, we note that in the case of K3-24 there is a dense torus. In this case we expect the companion mass to be $M_2 \ga 0.3 M_\odot$. 

We eject the dense (intensive) spherical wind that forms the dense shell starting one year after the end of the jet-launching episode, i.e., at $t=t_{\rm j}+1 \yr$, and continue with this mass loss until $t_{\rm w}=60 \yr$. We inject the dense wind at radius $r_{\rm w,in} = 4\times10^{14} \cm $.
The mass loss rate and velocity of the spherical dense wind are 
$\dot M_{\rm w} = 10^{-3} M_\odot \yr^{-1}$ and 
$v_{\rm w} = 20 \km \s^{-1}$, respectively. 
In one case we inject the regular AGB winds rather than a dense wind. 
The simulations of a dense shell that is younger than the jets, i.e., a post-jets shell, is the main new ingredient of our study with respect to our group's previous studies. From observations we know that jets can be younger or older than the dense shell that was presumably ejected in a common envelope evolution (e.g., \citealt{Tocknelletal2014}). In most cases the age difference between the jets and the dense shell is very small and we can refer to them as coeval  \citep{Guerreroetal2020}. 
   
We summarise the simulations we perform in Table \ref{Table:cases}. 
\begin{table*}
\centering
\begin{tabular}{|c|c|c|c|c|c|c|c|}
\hline

Simulation &  $\dot M_{\rm 2j}$          & $v_{\rm j}$  & $t_{\rm j}$ & $\alpha_{\rm j}$ & $\gamma$ & Figures & $\theta_{ears}$ \\ 
           &  $10^{-6} M_\odot \yr^{-1}$ & $\km \s^{-1}$& $\yrs$      &         &           &           &     \\ 
 \hline 
S1         & $38$ & $100$ & $1$ & $15^\circ$&$1.1$  & \ref{fig:sixteen_ears}, \ref{fig:3Gammas} & $40^\circ$ \\ \hline
S2         & $38$ & $100$ & $1$ & $15^\circ$ &$1.33$ &\ref{fig:sixteen_ears}, \ref{fig:3Gammas} & $35^\circ$ \\ \hline
S3         & $38$ & $100$ & $1$ & $15^\circ$ & $1.67$ & \ref{fig:sixteen_ears}, \ref{fig:3Gammas} & $25^\circ$\\ \hline
S4         &$9.5$ & $200$ & $1$ & $15^\circ$ &$1.1$ & \ref{fig:sixteen_ears}, \ref{fig:evolS4} & $35^\circ$ \\ \hline
S5         &$9.5$ & $200$ & $1$ & $15^\circ$ &$1.33$ & \ref{fig:sixteen_ears} & $30^\circ$\\ \hline
S6         &$9.5$ & $200$ & $1$ & $15^\circ$  &$1.67$ & \ref{fig:sixteen_ears}, \ref{fig:evolS6} & $30^\circ$ \\ \hline
S7         &$9.5$ & $200$ & $2$ & $15^\circ$ & $1.1$ & \ref{fig:sixteen_ears} & $30^\circ$ \\ \hline    
S8         &$152$ & $50$  & $1$ & $15^\circ$ &$1.67$  & \ref{fig:sixteen_ears} & $40^\circ$ \\ \hline
S9         &$9.5$ & $200$ & $1$ & $50^\circ$&$1.1$  & \ref{fig:sixteen_ears} & $40^\circ$ \\ \hline
S10        & $38$ & $100$ & $1$ & $50^\circ$ &$1.1$  & \ref{fig:sixteen_ears} & $35^\circ$ \\ \hline
S11        &$9.5$ & $200$ & $2$ & $15^\circ$ &$1.67$  & \ref{fig:sixteen_ears} & ($35^\circ$) \\ \hline
S12        &$9.5$ & $200$ & $3$ & $15^\circ$ & $1.67$  & \ref{fig:sixteen_ears} & ($10^\circ$)\\ \hline
S13        & $38$ & $100$ & $1$ & $50^\circ$ &$1.67$ & \ref{fig:sixteen_ears} & ($20^\circ$) \\ \hline
S14        & $38$ & $100$ & $1$ & $50^\circ$ &$1.33$  & \ref{fig:sixteen_ears} & ($20^\circ$) \\ \hline
S15        & $9.5$ & $200$ & $1$ & $50^\circ$ &$1.33$ & \ref{fig:sixteen_ears} & ($40^\circ$) \\ \hline
S16        & $9.5$  & $200$  & $1$ & $50^\circ$ &$1.67$  & \ref{fig:sixteen_ears} & ($30^\circ$) \\ \hline
S1L        & $38$ & $100$ & $1$ & $15^\circ$ & $1.1$  & \ref{fig:S1L} & $40^\circ$ \\ \hline
\end{tabular}
\caption{Summary of the 17 simulations we present in the paper. The columns list, from left to right and for each simulation, its number, the mass loss rate of the two jets combined $\dot M_{\rm 2j}$, the velocity of the jets $v_{\rm j}$, the time period of jets' activity $t_{\rm j}$, the half opening angle of the jets $\alpha_{\rm j}$, and the adiabatic index $\gamma$.
In the next column we list the figures presenting each simulation.
 In all figures beside Fig. \ref{fig:incliden} that we present later on, the symmetry axis is on the plane of the sky, i.e., $i=90^\circ$. In the last column we list the critical inclination angle $i_{\rm ears}$ (defined as the angle between the PN symmetry axis and the line of sight) for each case, below which the ears disappear because they are projected on the main shell.  In all cases we start at $t=0$ with a regular AGB wind that fills the grid, and we start to inject the dense shell one year after the end of the jets' activity, i.e., at $t=t_{\rm j} + 1 \yr$. In Simulation S1L we inject a regular AGB wind instead of a dense wind during the post-jets phase.  
}
\label{Table:cases}
\end{table*}

\section{RESULTS}
\label{sec:results}

\subsection{A gallery of images}
\label{subsec:gallery}
    
\begin{figure*}[ht!]
\includegraphics[trim=0.6cm 10.2cm 0.0cm 2.4cm ,clip, scale=0.95]{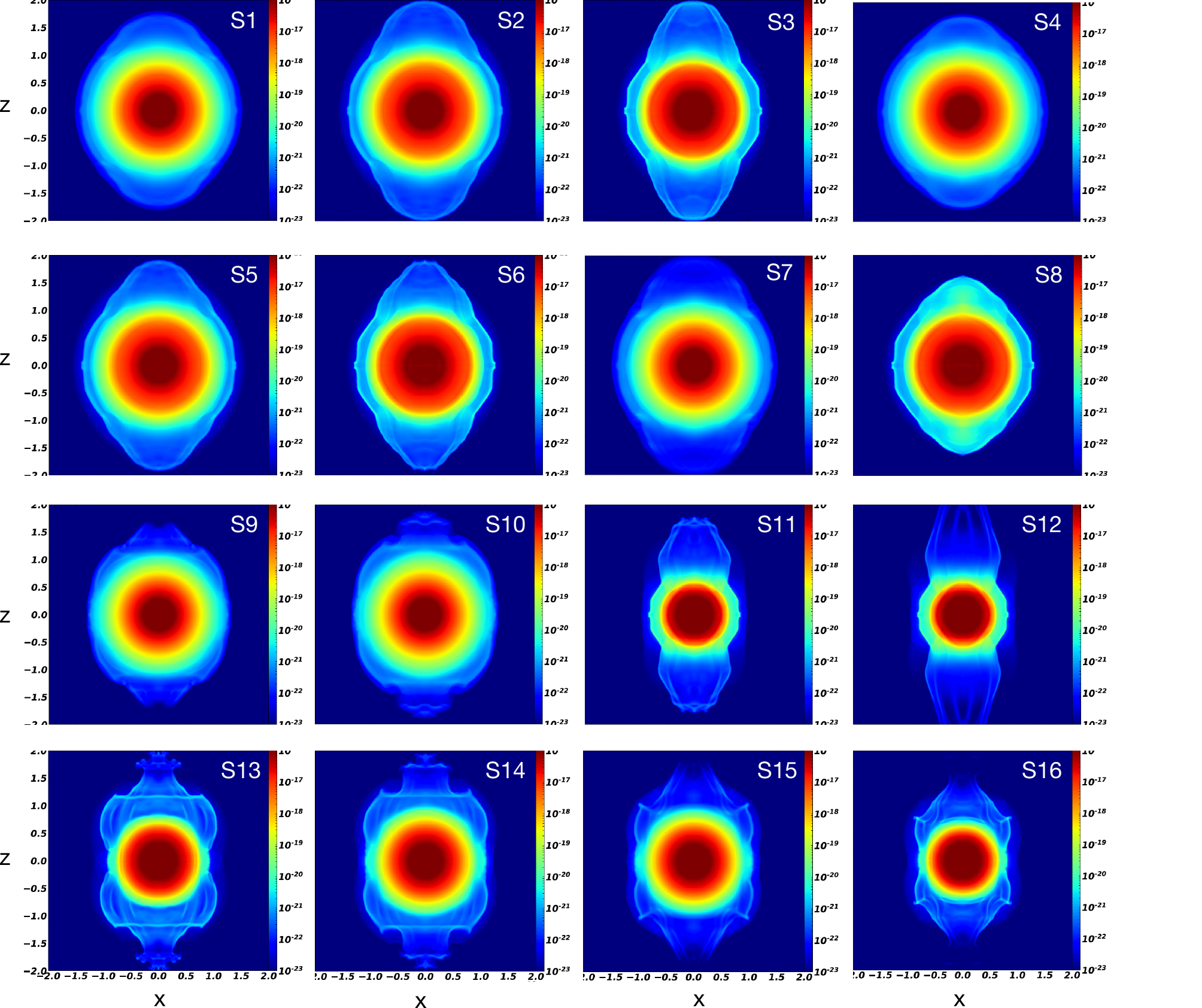} \\
\caption{Artificial intensity maps for 16 models. Each artificial intensity map is a map of the integration of density squared along the $y$ axis (the line of sight). In all cases the symmetry axis of the two opposite jets is $(x,y,)=(0,0)$, namely, through the center and along the $z$ axis. All panels are square with sizes of $4 \times 10^{16} \cm$. The colors depict the artificial intensity values according to the color bars in the range  of $10^{-23} \g^2 \cm^{-5} - 10^{-16} \g^2 \cm^{-5}$. We consider simulations S1 to S7 to yield ears, simulations S9, S15 and S16 to be marginal, and simulations S10-S14 to yield no ears. }
	\label{fig:sixteen_ears}
\end{figure*}

We start by comparing 16 simulations that we performed when the bipolar structure reach about the same size as each other. In Fig. \ref{fig:sixteen_ears} we present the artificial intensity maps of these 16 cases. The artificial intensity map is a map of the integration of density squared along the line of sight, here along the $y$ axis. In all simulations we start to blow the dense shell a year after we turned off the jets. Namely, the jets are older than the main nebular shell. For other properties see Table \ref{Table:cases}.  
    
We recall our definition of ears as two opposite protrusions from the main shell that (1) are smaller than the main inner shell from which they protrude, (2) have a cross section (perpendicular to the symmetry axis) that monotonically decreases outward, and  (3) the boundary between the ears and the main nebula has a dimple (two inflection points) on each side of each ear.   
We clearly identify ears in simulations S1 to S7, but in simulation S7 the ears are almost too large and very faint. Cases S9, S15,  and S16 are marginal as the ears do not have a clear shape as in simulations S1-S7, and are fainter. In cases S10-S14 we do not identify the faint protrusions as ears. 
  
Our first conclusion is that the flow sequence of weak jets that interact with a regular AGB wind followed by the ejection of a dense shell (an intensive wind) can lead to ear formation, but not necessarily so. 

\subsection{The role of the adiabatic index}
\label{subsec:Adiabatic}

\begin{figure*}[ht!]
\includegraphics[trim=0.8cm 8.1cm 0.0cm 2.0cm ,clip, scale=0.90]{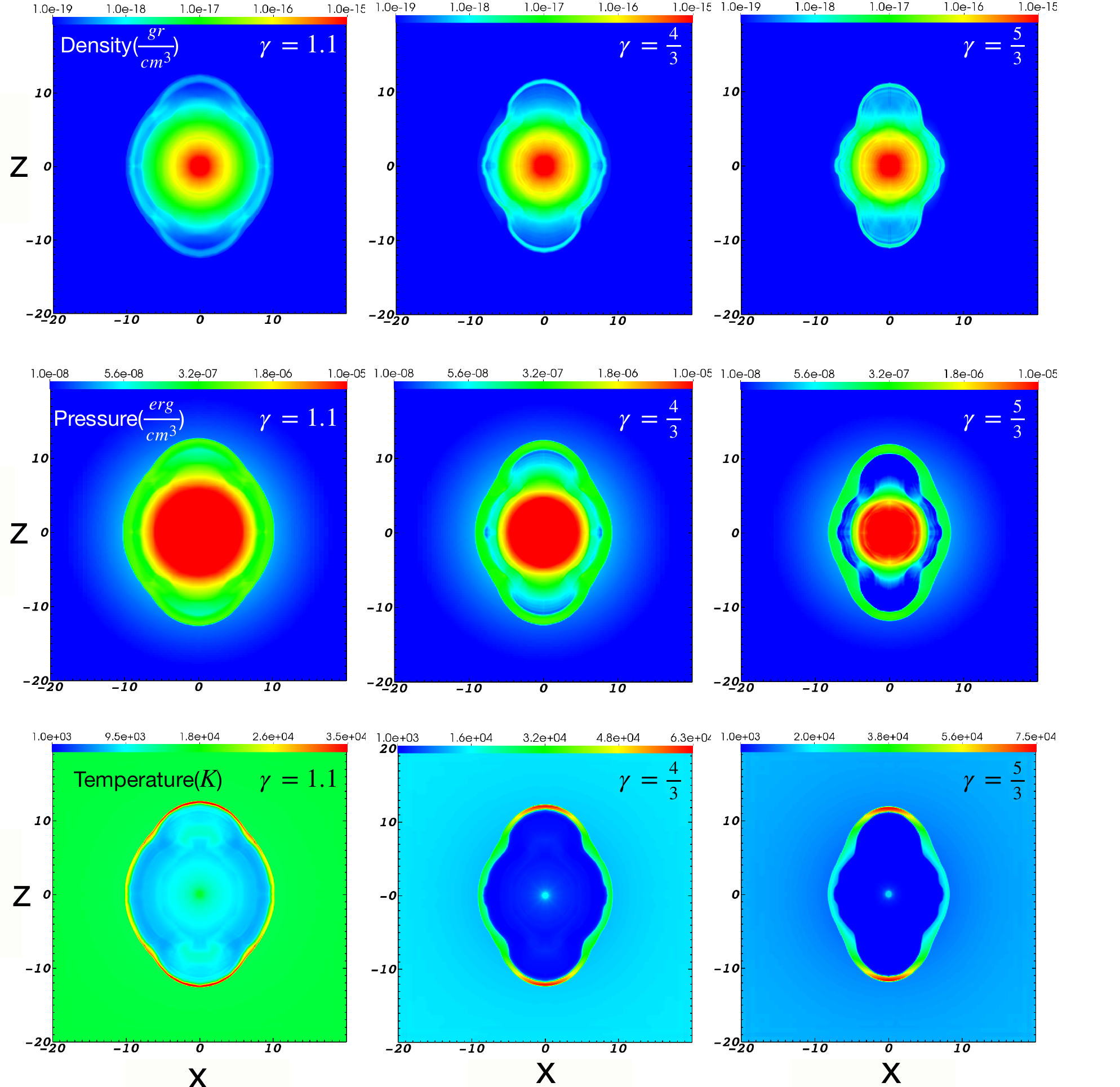} \\
	\caption{Comparing the density (upper row), pressure (middle row), and temperature (lower row) maps in the meridional plane of three simulations that differ only in the value of the adiabatic index. Left column: simulation S1 with $\gamma=1.1$; Middle column: simulations S2 with $\gamma=1.333$; Right column: simulation S3 with $\gamma=1.67$. All panels are square with sizes of $4 \times 10^{16} \cm$ and at $t=44 \yr$. The numbers on the axis are in units of $10^{15}\cm$.   Densities according to color-bars in the range of $10^{-19} \g \cm^{-3}- 10^{-15} \g \cm^{-3}$, while pressure in the range of $10^{-8} \erg \cm^{-3}- 10^{-5} \erg \cm^{-3}$. Note that the densities in the zone $r \ge10 \times 10^{15} \cm$ are $\le 2.5 \times 10^{-20} \g \cm^{-3}$ (decreasing as $r^{-2}$) and so appear all blue. The temperature ranges are from $1000 \K$ (blue) to $3.5\times 10^4 \K$ in the lower-left panel, to $6.3\times 10^4 \K$ in the lower-middle panel, and to $7.5\times 10^4 \K$ in the lower-right panel. } 
	\label{fig:3Gammas}
\end{figure*}

In Fig. \ref{fig:3Gammas} we compare the density, pressure, and temperature maps in the meridional plane $y=0$ of simulations S1, S2, and S3 (from left to right) that differ only by the value of the adiabatic index $\gamma$. 

The adiabatic index plays a role in both increasing and decreasing the temperature. A higher value of $\gamma$ implies a steeper change in pressure as density changes. In these three simulations the jets start highly supersonic, with a mach number of $\mathcal{M}_{\rm j} = 6.7$. In the postshock region the Mach number and temperature increase as $\gamma$ increases. Indeed, in the lower three panels of Fig. \ref{fig:3Gammas} we see that the higher the value of $\gamma$ is the higher the temperature of the post-shock jets' gas is (note that the red color stands for a higher temperature as $\gamma$ increases in the three panels). On the other hand, as the gas expands a higher value of $\gamma$ implies more rapid loss of pressure; this reduces the expansion velocity.
For example, in a gas that is set to expand freely into an empty tube the maximum velocity at the front of the expanding gas is $2 C_{0} / (\gamma-1)$, where $C_0$ is the initial sound speed of the gas. Namely, the maximum additional velocity of the expanding gas is proportional to $(\gamma-1)^{-1}$.  
In simulations where the jets are active for a long time, the effect of higher post-shock pressures for higher values of $\gamma$ dominates, and flow with higher values of $\gamma$ inflates larger bubbles. The present flow structure has short-lived and weak jets and a slow pre-jet wind and a slow post-jet intensive wind (the dense shell), i.e., a Mach number of only $\mathcal{M}_{\rm s} = 1.3$ for both winds. The result is that  the effect of a faster cooling for higher values of $\gamma$ dominates in many parts. 
Indeed, we see that the high-pressure region (red color in middle row of Fig. \ref{fig:3Gammas}) gets smaller as $\gamma$ increases, and that the temperature in the center is the highest for the lowest value of $\gamma$. As well, the hot thin shell is larger for the lower values of $\gamma=1.1$ and smaller for $\gamma=1.67$, in particular in the equatorial plane. 
 
Another comparison is of simulation S7 and S11. In these two simulations the jets have the same power as in simulations S1-S6, but the jets are active for $t_{\rm j}=2 \yr$ instead of for only $t_{\rm j}=1 \yr$. Namely, the jets deposit twice as much energy to the lobes/ears they inflate with respect to simulations S1-S6 (we discuss this further in section \ref{subsec:EnergyMomentum}). In simulation S11 that has a larger value of $\gamma=1.67$ the jets inflate narrower lobes that form a bipolar PN rather than ears. These lobes are not ears because the cross section does not decrease monotonically as we move out. In simulation S7 for which $\gamma=1.1$ the lobes are wide, and almost larger than the dense shell. These are nonetheless ears.  

\subsection{The role of energy and momentum of the jets}
\label{subsec:EnergyMomentum}

There are five pairs and one triplet of simulations with the same adiabatic index $\gamma$ and the same power and duration of jets, but different momentum. The pairs are (S1,S4), (S2,S5), (S10,S9), (S13,S16), and (S14,S15), where the first simulation in each pair is the one with twice as large momentum flux compared with the second simulation in the pair. Overall, in the simulations with higher jets' momentum, all other parameters being similar, the lobes/ears are more elongated. As well, in the marginal cases (S10,S9) and (S13,S16) the higher momentum forms a wider lobe/ear on the far zone (far from the center), and therefore the cross section of the lobe/ear does not decrease monotonically. This prevents the lobes from being defined as ears.  

In the triple (S8,S3,S6) the jets in simulation S8 have twice the momentum of that in simulation S3, that in turn has twice the jets' momentum in simulation S6. While in simulations S3 and S6 we do obtain ears, in simulation S8 the jets' velocity of $v_{\rm j} =50 \km \s^{-1}$ is too low for the jets to inflate ears or lobes and we obtain an elliptical nebula. 

Simulations S7 and S11 are active for twice as long, while simulation S12 is active for three times as long as the other simulations. In these simulations, in particular S11 and S12, the jets inflate too large lobes to be defined as ears. As expected, energetic jets form bipolar nebulae.

\subsection{The role of jets' opening angle}
\label{subsec:Angle}

There are simulations where we inject wide jets with a half opening angle of $\alpha_{\rm j}=50^\circ$ instead of $\alpha_{\rm j}=15^\circ$.
Pairs with narrow and wide, in this order, jets but otherwise identical simulations are (S1, S10), (S4,S9), (S3,S13), (S2,S14), (S5,S15), and (S6,S16).  We learn from these comparisons that too wide jets form complicated faint structures in the polar direction that are not what we refer to as ears. 
Basically, the wide jets inflate large lobes, which because of instabilities form a bumpy outer boundary of the ear, as well as a cross section that not always decreases monotonically outward. These effects prevent the inflated lobe to obey our definition of ears. 

\subsection{The appearance of ears}
\label{subsec:appearance}
 
The interaction of the jets with the regular pre-jets AGB wind forms the ears. The post-jets dense wind forms the dense nebular shell, but otherwise plays no hydrodynamical role in forming the ears. The dense shell serves to form nebulae similar to most of the observed PNe with ears, where the ears are fainter than the main nebula. If there is no dense wind but rather the regular AGB wind continues in the post-jets phase, the ears might merge with the nebula to form an elliptical PN without ears. We demonstrate this with simulation S1L.  

\begin{figure*}[ht!]
\includegraphics[trim=0.8cm 0.1cm 0.0cm 0.0cm ,clip, scale=0.20]{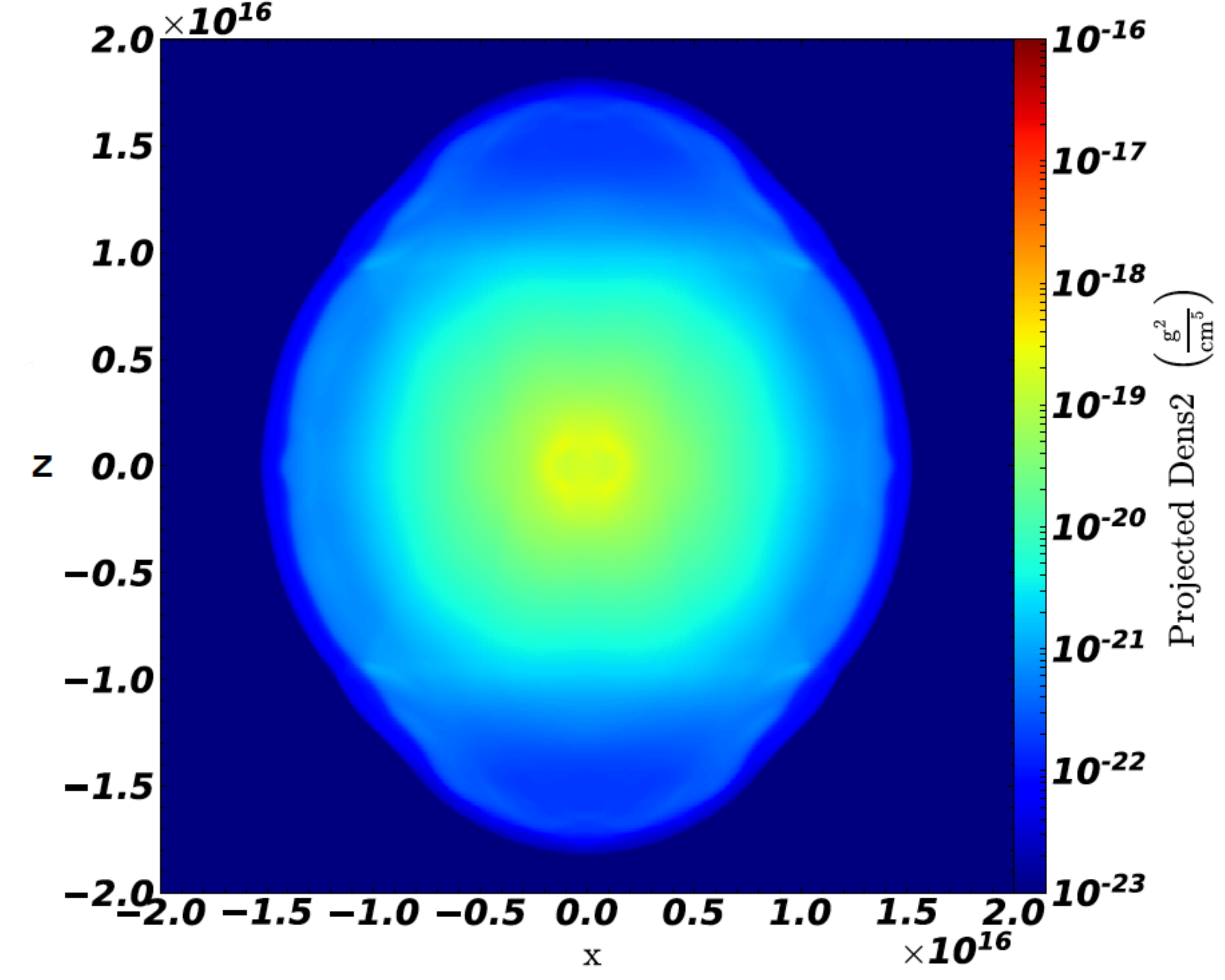} \\
\caption{ The artificial intensity maps of simulation S1L that has the same parameters as simulation S1 but with a regular AGB wind in the post-jets phase (Table \ref{Table:cases}). The time is the same time as in the upper left panel of Fig. \ref{fig:sixteen_ears} for simulation S1.	} 
	\label{fig:S1L}
\end{figure*}

In simulation S1L the jets properties are as in simulation S1, but instead of injecting a post-jets dense wind, i.e., with a mass loss rate and velocity of 
$\dot M_{\rm w} = 10^{-3} M_\odot \yr^{-1}$ and $v_{\rm w} = 20 \km \s^{-1}$, respectively, we inject the regular AGB wind with a mass loss rate and velocity of 
$\dot M_{\rm AGB} = 10^{-6} M_\odot \yr^{-1}$ and 
$v_{\rm AGB} = 20 \km \s^{-1}$, respectively. We present the artificial intensity maps of simulation S1L in Fig. \ref{fig:S1L}. We do indeed form ears. However these ears are only marginally fainter than the main nebular outskirts, whereas in simulation S1 (upper left panel of Fig. \ref{fig:sixteen_ears}) the ears are much fainter than the nebula. We suspect that in the case of simulation S1L the ears will merge with the main nebula at a later phase after ionisation starts, and will form an elliptical PN without ears. 
In summary, to form ears that are fainter than the nebula (or nebula brighter than the ears) we find that we should increase the wind mass loss rate in the post-jets phase.    

\subsection{The critical angle for ears}
\label{subsec:critical}

\begin{figure*}[ht!]
\includegraphics[trim=0.9cm 11.3cm 0.0cm 1.0cm ,clip, scale=0.8]{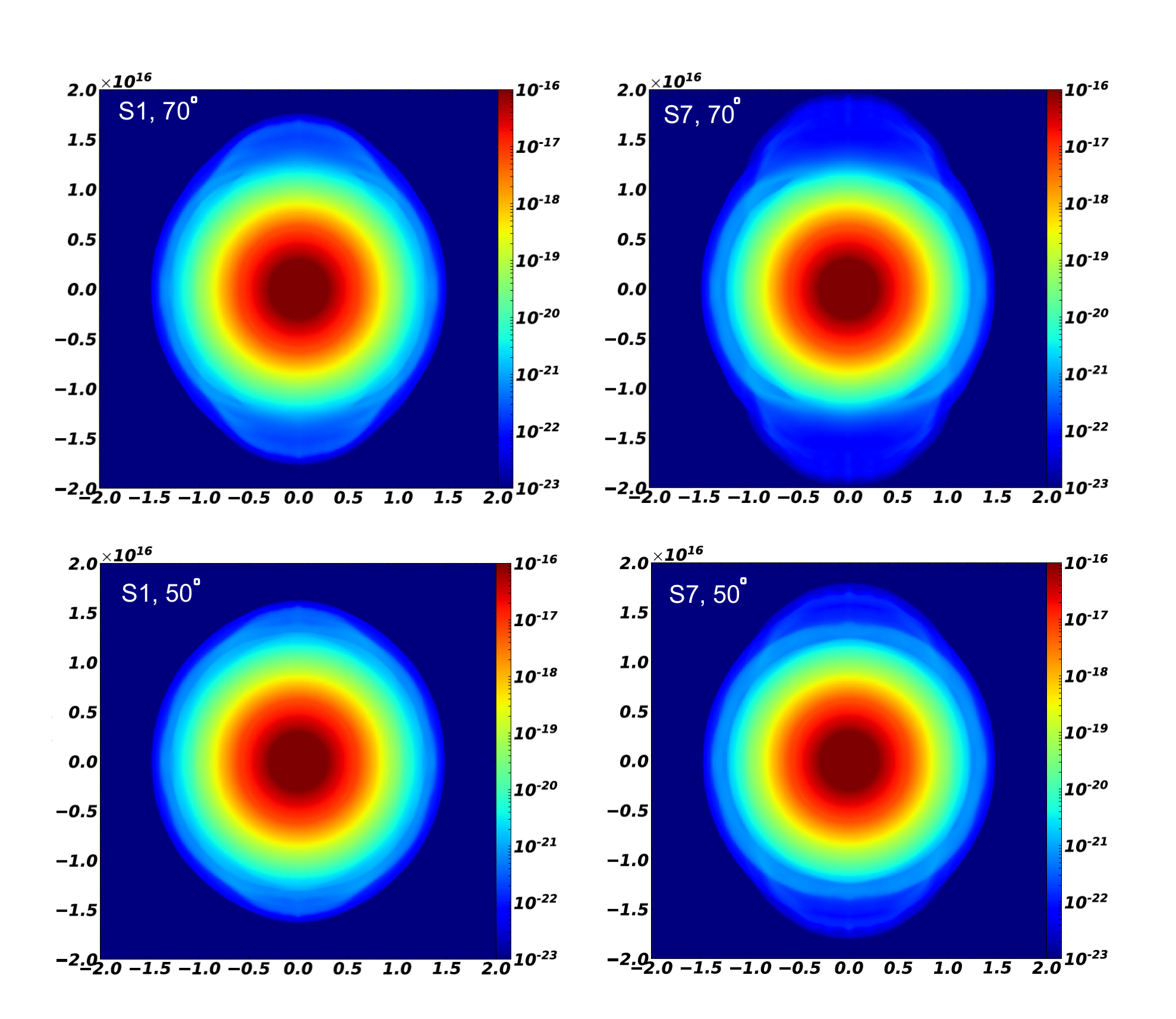} \\
\caption{ Artificial intensity maps for simulations S1 (left column) and S7 (right column) and for two inclination angles (the angle between the symmetry axis of the PN and the line of sight). Each artificial intensity map is a map of the integration of density squared along the line of sight for an inclination angle $i$ as indicated in the inset. The ears disappear as the inclination angle decreases.
} 
	\label{fig:incliden}
\end{figure*}

Finally we refer to the inclination angle. In all figures beside Fig. \ref{fig:incliden} we present the images with an inclination angle of $i=90^\circ$, i.e., the symmetry axis of the PN (through the two ears) is in the plane of the sky.   In Fig. \ref{fig:incliden} we present the artificial intensity maps of two cases and at two inclination angles as we indicate in the four panels. These demonstrate how the ears becomes less prominent as the inclination angle decreases. 

   For small inclination angles of $i < i_{\rm ears}$ the ears are projected on the main nebula and we cannot notice them by the morphology. For each simulation we examine at what inclination angle, the critical inclination angle $i_{\rm ears}$, the ears disappear at the end of the simualtion. Namely, we can observe ears only for $i>i_{\rm ears}$. We list these values (to accuracy of $5^\circ$) in the last column of Table \ref{Table:cases}. We list the critical angle also for cases where we see no ears, cases where the angle is inside parenthesis. In these cases the angle is for the disappearance of the polar protrusions even if they are not ears. Because in all our simulations $i_{\rm ears} \la 35^\circ -40^\circ$, a random orientation of the PN symmetry axis implies that we miss ears because of projection on the main PN shell only in $\simeq 20 \%$ of the cases.

\section{Evolution}
\label{sec:Evolution}

\begin{figure*}[ht!]
\includegraphics[trim=0.9cm 2.3cm 0.0cm 1.0cm ,clip, scale=0.8]{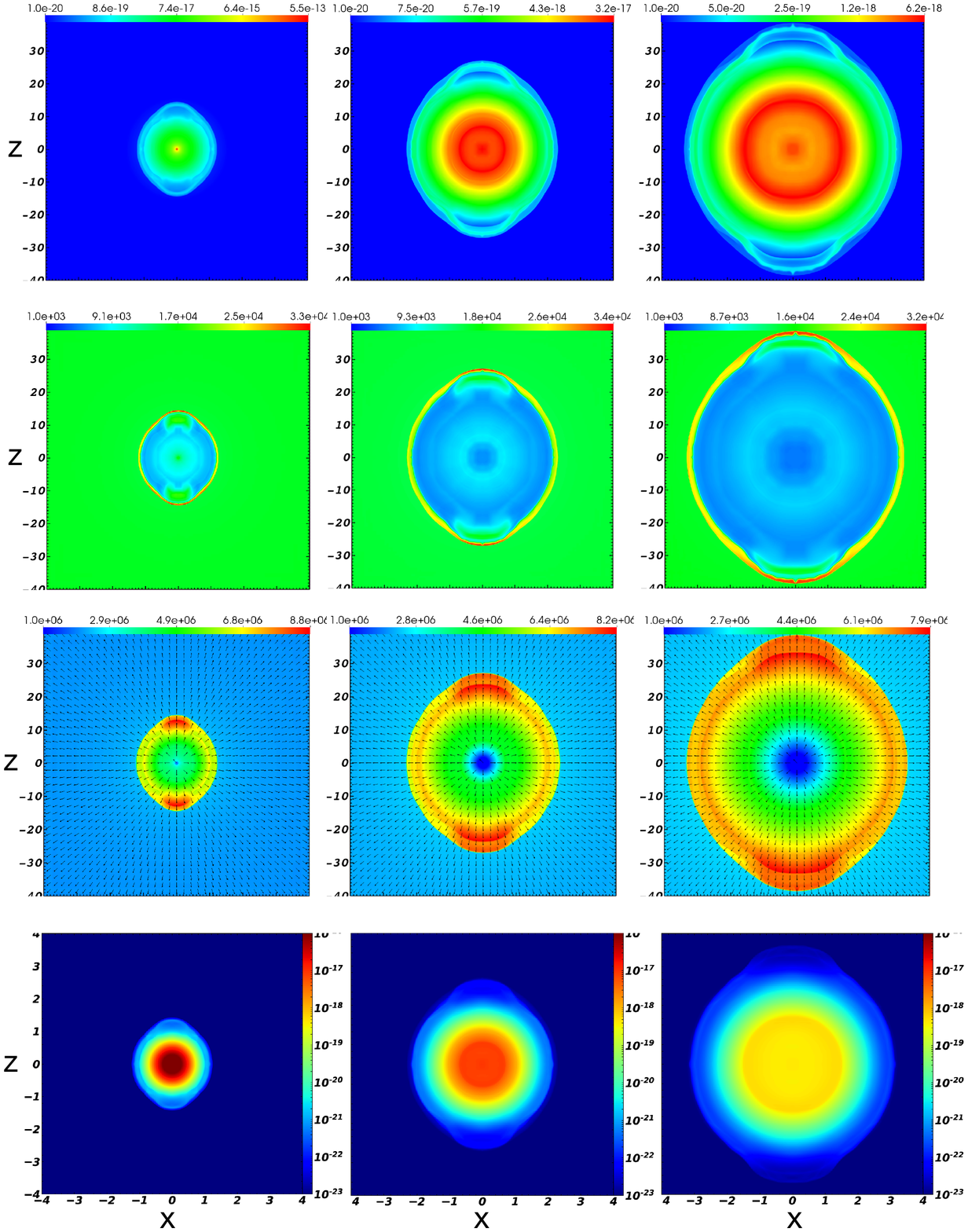} \\
\caption{Evolution of simulation S4 at three times, from left to right, $t=54 \yr$, $t=104 \yr$ and $t=152\yr$. We present the density (upper row), temperature (second row), and velocity magnitude according to the colors with arrows indicating the flow direction (third row), all in the meridional plane $y=0$ and with the color-bars in cgs units. In the lower row we present the evolution of the artificial intensity map (in units of $\g^2 \cm^{-5}$ according to the color-bar), where the first panel is as in Fig. \ref{fig:sixteen_ears}. } 
	\label{fig:evolS4}
\end{figure*}

We present the evolution of two simulations.
In Fig. \ref{fig:evolS4} we present, from top to bottom, the density, the temperature, and the velocity map in the meridional plane $y=0$ of simulation S4 at three times, from left to right. In the bottom row we present the artificial intensity map (integration of density squared along the line of sight, here along $y$). As we observe at $t=152 \yr$, when the ears reach the edge of the grid, the ears maintain their identity. As the entire nebula is supersonic, Mach numbers $\mathcal{M} > 3$ in most parts, and most of the motion is radial, the nebula will keep its structure at later times as well (unless a too massive circumstellar material further out will change that structure). 
This simulation shows that for some physical parameters the ears can exist for hundreds of years and more. 

\begin{figure*}[ht!]
\includegraphics[trim=1.0cm 3.cm 0.0cm 0.0cm ,clip, scale=0.8]{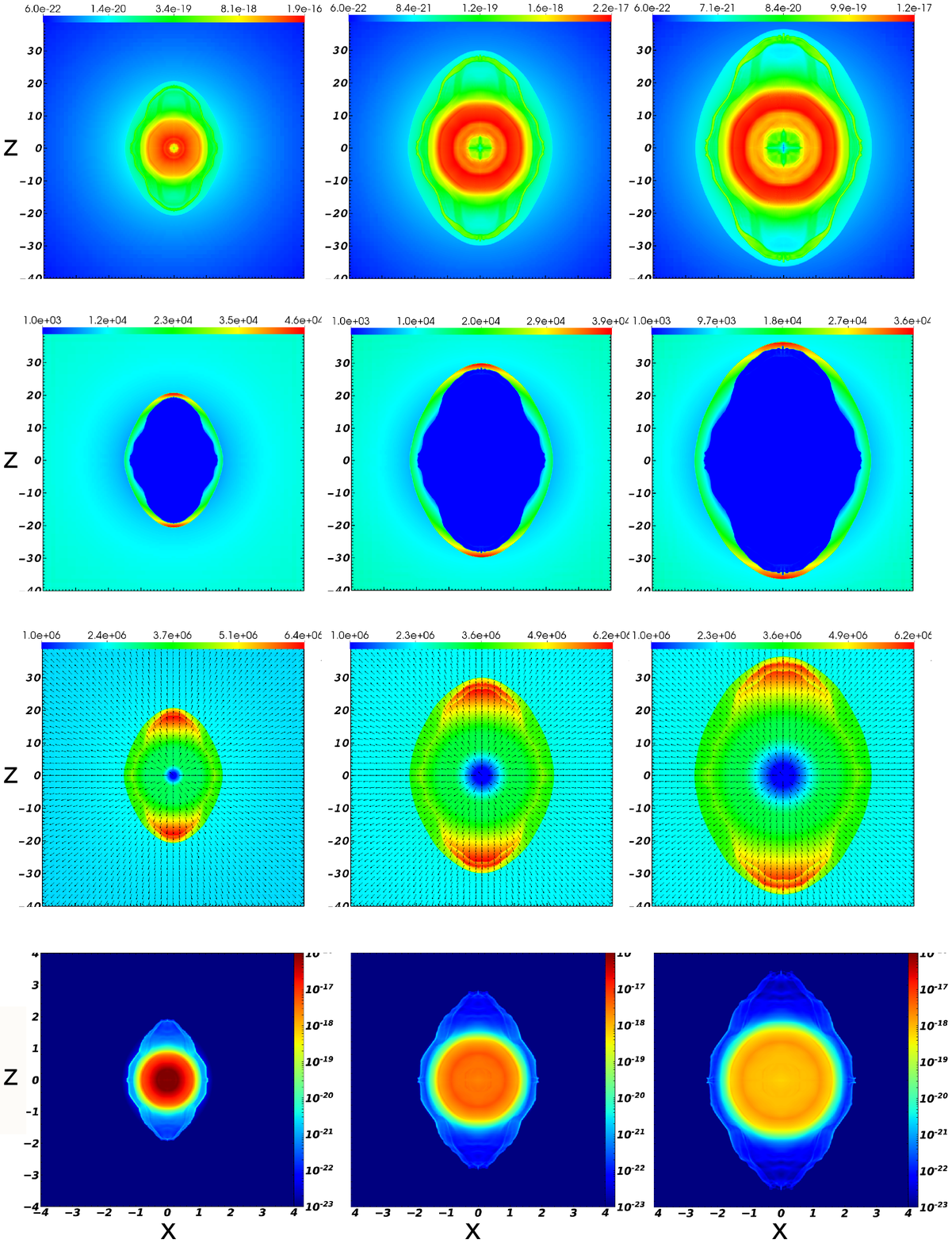} \\
\caption{Similar to Fig. \ref{fig:evolS4} but for simulation S6 and at the three times of $88 \yr$, $136 \yr$ and $171\yr$.  	} 
	\label{fig:evolS6}
\end{figure*}

In Fig. \ref{fig:evolS6} we present the evolution of simulation S6. The same discussion above for simulation S4 holds for this case as well. Basically, although our simulations in both S4 and S6 are for less than 200 years, at the end of the simulation the flow is radial and supersonic, and we expect the ears morphological feature to stay for thousands of years. 

Saying all these, we did not follow the nebula to the phases when the central star blows a fast ($\gg 100 \km \s^{-1}$) wind and starts to ionise the nebula. The fast wind interaction with the dense shell influences the evolution at later times (e.g., \citealt{Perinottoetal2004}), e.g., it suffers from instabilities and destroys the smooth structure of the dense shell (e.g., \citealt{ToalaArthur2016}). We expect that the dense shell will nonetheless contain the fast wind such that the fast wind will not affect the ears. The ionisation of the nebula will increase the sound speed and somewhat will change the flow (e.g., \citealt{Perinottoetal2004, Schonberneretal2010}). This might erase small and faint ears, or smear the differences between the ears and the main nebula such as in simulation S1L (Fig. \ref{fig:S1L}), but in most cases we expect ears to survive these late evolutionary phases. Future simulations should examine the role of the fast wind and the ionising radiation to examine whether the ears survive as we expect. 

\section{Summary}
\label{sec:summary}
 
The morphologies of a small fraction of elliptical PNe contain two opposite protrusions from the main PN shell that are smaller than the main PN shell, have a cross section that decreases monotonically outward, and  the boundary between the ears and the main nebula has a dimple (two inflection points) on each side of each ear.   These two opposite protrusions are termed `ears' (examples are in section \ref{sec:intro}). Our goal was to determine the outflow structure by which jets can inflate ears. In many trials that we do not present here, we could not obtain ears when we launched the jets after we blew the main dense shell. Namely, the jets that interact with the dense shell either do not inflate any protrusions, or if they do inflate protrusions these are large lobes that form bipolar PNe. 

We therefore simulated here a flow structure where low-energy jets (short-lived and not too powerful) interact with a regular AGB wind, and the dense PN shell is younger than the jets (for details see section \ref{sec:numerical}). In these simulations, that we summarise in Table \ref{Table:cases},  we started to blow the intensive wind that forms the dense inner shell one year after the jets ceased. We assumed that the main sequence companion is of low mass, $M_2 \simeq 0.1-0.3 M_\odot$ (might even be a brown dwarf), and for that it launches weak jets that form the ears, and after it enters the CEE it ejects an elliptical nebula rather than a bipolar nebula or a dense equatorial torus. This assumption is compatible with the presence of ears only in elliptical PNe. In many cases we expect that the low mass companion will not survive the CEE (it will spiral-in all the way to the core and be tidally destroyed). Even if it survives, its low mass implies that it is hard to detect such companions in the centres of PNe. We referred to the PN A30 as an example of an elliptical PN with a post-CEE central binary system \citep{Jacobyetal2020}. 
 
The full parameter space is huge as we can vary the jets' opening angle, the mass loss rate into the jets and their velocity, the properties of the regular AGB wind into which the jets expand, and the adiabatic index. For the influence of the adiabatic index see Fig. \ref{fig:3Gammas}. Indeed, from the 16 simulations we conducted we identify clear ears in seven, S1-S7 in Fig. \ref{fig:sixteen_ears}. 
 
We found that not under all conditions we form ears. We found that the jets cannot be too energetic, cannot be too wide, and cannot be too slow. At the end of our simulations the outflow is radial and supersonic, and so the jets maintain their morphology for hundreds of years (section \ref{sec:Evolution}; Figs. \ref{fig:evolS4}, \ref{fig:evolS6}), and probably much longer.  

Our main finding is that weak and short-lived jets that a companion launches before it enters the CEE might form ears in elliptical PNe. We can present this from another perspective where we refer to the large jets' parameter space. Namely, jets that are weak, short-lived, and launched before the main nebular ejection, lead to the formation of ears in elliptical PNe.    

Because the parameter space is too large to follow in one study, there is much more studies to do before we can clearly reproduce specific PNe with ears. For example, we should conduct 3D hydrodynamical simulations of a binary system that launches jets as it enters a CEE, similar to the simulation by \cite{Shiberetal2019}.
As well, we should continue the simulations for thousands of year and include the central fast wind and the ionisation phase of the PN. However, we think that we can confidently state that to form PNe with ears,  in the binary-jet paradigm, the progenitor binary system should launch the jets shortly before it blows the dense PN shell. 
Such a flow structure can come from a system that enters a common envelope evolution. The companion accretes mass through an accretion disk just before it enters the envelope of the AGB star (or even a red giant branch star), and launches jets for a short time. It then enters the envelope and ejects the envelope to form the dense shell of the descendant PN. In other words, our results are consistent with a scenario in the frame of the binary-jet paradigm where in PNe with ears, the progenitor binary system launched the jets shortly before the system entered the common envelope evolution. 

We do not claim that the binary-jet scenario is the only one to form ears. { For example, the two lobes of a bipolar PN with a small inclination angle,  defined as the angle between the PN symmetry axis and the line of sight, (i.e., an almost pole-on PN) might appear as two ears protruding from the main nebula. } It does, however, have some expectations that we find in some PNe, and so might support this scenario. The binary-jets scenario includes the possibility that in some cases the accretion disk will precess, and so will the jets that it launches. As well, in some cases, mainly due to a more massive companion, there will be a dense equatorial outflow during the CEE phase. 
The PN K3-24 that we list in section \ref{sec:intro} has two pairs of ears not aligned perpendicular to the dense equatorial gas. This clearly suggests a binary interaction. The "S" shape of the ears both in K3-4 and in NGC~6563 suggests precession, which in turn suggests binary interaction.  
  
\section*{Acknowledgments}

 We thank an anonymous referee for very useful and detailed comments.  
This research was supported by a grant from the Israel Science Foundation (769/20) and a grant from Prof. Amnon Pazy Research Foundation.



\label{lastpage}
\end{document}